\documentclass[conference]{IEEEtran}
\IEEEoverridecommandlockouts
\usepackage{cite}
\usepackage[numbers,sort&compress]{natbib}

\usepackage[utf8]{inputenc}
\usepackage{amsmath,amssymb}
\usepackage{graphicx}
\usepackage{bm}
\usepackage{enumerate}
\usepackage{comment}
\usepackage{xcolor}
\usepackage{url}
\usepackage{color,soul}
\usepackage{subcaption}
\usepackage{float}

\definecolor{light-gray}{gray}{0.8}

\usepackage{amsmath,amssymb,amsfonts}
\usepackage{algorithmic}
\usepackage[linesnumbered,ruled]{algorithm2e}
\usepackage{graphicx}
\usepackage{textcomp}
\usepackage{xcolor}
\usepackage{subcaption}

\def\BibTeX{{\rm B\kern-.05em{\sc i\kern-.025em b}\kern-.08em
    T\kern-.1667em\lower.7ex\hbox{E}\kern-.125emX}}

\makeatletter
\newcommand{\linebreakand}{%
  \end{@IEEEauthorhalign}
  \hfill\mbox{}\par
  \mbox{}\hfill\begin{@IEEEauthorhalign}
}
\makeatother

\begin{document}

\title{Cloud-Based AI Systems: Leveraging Large Language Models for Intelligent Fault Detection and Autonomous Self-Healing\\}

\author{

\small 

\begin{tabular}[t]{c@{\extracolsep{8em}}c} 

\textsuperscript{} Cheng Ji & Huaiying Luo\\
\textsuperscript{} Siebel School of Computing and Data Science & College of Computing and Information Science \\
\textit{University of Illinois Urbana-Champaign} & \textit{Cornell University}\\
\textsuperscript{}Champaign, Illinois, USA & New York, USA \\
\textsuperscript{}chengji5@illinois.edu & hl2446@cornell.edu \\

\\

\end{tabular}
}

\maketitle

\begin{abstract}
With the rapid development of cloud computing systems and the increasing complexity of their infrastructure, intelligent mechanisms to detect and mitigate failures in real time are becoming increasingly important. Traditional methods of failure detection are often difficult to cope with the scale and dynamics of modern cloud environments. In this study, we propose a novel AI framework based on Massive Language Model (LLM) for intelligent fault detection and self-healing mechanisms in cloud systems. The model combines existing machine learning fault detection algorithms with LLM's natural language understanding capabilities to process and parse system logs, error reports, and real-time data streams through semantic context. The method adopts a multi-level architecture, combined with supervised learning for fault classification and unsupervised learning for anomaly detection, so that the system can predict potential failures before they occur and automatically trigger the self-healing mechanism. Experimental results show that the proposed model is significantly better than the traditional fault detection system in terms of fault detection accuracy, system downtime reduction and recovery speed. 
\end{abstract}


\section{Introduction}
The Cloud is the backbone of modern computing resource architecture and draws it all together, from variants responsible for enterprise resource management, right through to applications supporting AI powered service. Such systems can offer scalability, flexibility and efficiency, and are the engine that drive many technologies today \cite{li2025advancesappflcomprehensiveextensible}. However, the fast evolution and greater complexity of these cloud environments also poses challenges, especially pertaining to keeping the system up and running and minimizing downtime.

As cloud systems become larger and larger, the risk of faults such as hardware failures, network outages, and software defects increases which makes detecting faulty components and removing them through a self-healing mechanism a critical aspect of running a fault-tolerant cloud system.

In traditional cloud systems, fault detection often relies on pre-configured rules or human intervention. While these approaches are effective to some extent, they are often inadequate when it comes to dealing with the dynamics and diversity of modern cloud environments \cite{yang2025researchlargelanguagemodel}. As system architectures become more complex and the amount of data generated increases dramatically, traditional approaches are struggling to keep up with the speed of system change. As a result, there is an urgent need for more intelligent and adaptable systems that can automatically detect, diagnose, and correct faults on a real-time basis without human intervention.

Against this backdrop, artificial intelligence (AI) shows great potential in different area, especially when combined with advanced technologies such as machine learning (ML) \cite{jordan2015machine, mahesh2020machine, li2024exploring, yang2025data, he2023t, he2024give, jin2024scam, liu2024mt2st} and natural language processing (NLP) \cite{chowdhary2020natural, wu2021deep, li2024deception, 10628639, ji-etal-2024-rag}. Machine learning models can analyze large amounts of data generated by cloud systems to identify patterns and anomalies that may indicate failure \cite{jin2025adaptivefaulttolerancemechanisms}.

Furthermore, with the advent of large-scale language models (LLMs), new opportunities have been created to improve fault detection capabilities. LLMs are well-suited for processing and understanding unstructured data such as system logs, error messages, and user feedback \cite{he2025intent, rajput2021}, which is not possible with traditional models. Given its applications across diverse domains such as medical treatment, music composition, code generation, and narrative creation \cite{liu2025memeblip2, ding2024enhance, deng2024composerx, Ji2025, li2025visual, 10.1145/3627673.3679071}, it is reasonable to anticipate its effectiveness in the domain of fault detection of cloud system as well.

In addition to fault detection, the ability to automatically repair or mitigate failures is critical to ensuring high system availability and reducing operational overhead. Self-healing systems are designed to automatically detect problems and take corrective action, such as restarting services, reallocating resources, or applying patches, without the need for human intervention \cite{yang2024hades, jin2025scalability}. However, designing an effective self-healing system requires a robust framework that not only detects failures, but also predicts potential problems before they lead to longer downtimes.

\section{Related Work}
Lakshmi et al. \cite{lakshmi2023} proposed a self-healing network architecture that uses intelligent protocols to achieve autonomous fault detection and recovery. The study emphasizes intelligence at the protocol level, which can automatically identify and repair faults at the network level, improving the reliability and stability of the network. 

Ravi et al. \cite{vankayalapati2022} propose an AI-driven, self-healing cloud infrastructure framework that aims to enable autonomous recovery from sudden runtime failures. The framework uses deep reinforcement learning and other technologies to realize automatic fault detection, inference-based fault diagnosis, and accelerate the repair process.

Gireesh \cite{kambala2024} proposes a distributed software system architecture for mission-critical applications that integrates intelligent fault detection and self-healing mechanisms. The architecture combines the methods of artificial intelligence and software engineering to improve the robustness and adaptability of the system in the face of complex environments.

Devi et al \cite{devi2024} proposed a self-healing IoT sensor network architecture based on the Isolation Forest algorithm to realize autonomous fault detection and recovery. This method uses unsupervised learning technology to identify abnormal patterns and take corresponding self-healing measures to improve the reliability and efficiency of the network.

Tan et al. \cite{tan2021} point out that self-healing materials can be automatically repaired in robotic components, resulting in longer service life, lower maintenance costs, and reduced e-waste. A variety of self-healing mechanisms are also discussed, including microencapsulation-based chemical remediation, heat-activated shape memory alloys, and ion-crosslinked-based hydrogels.

Bao et al. \cite{bao2024} used a distributed smart distribution network, combined with fault location, load calculation, and self-healing control technologies, to achieve rapid fault detection and automatic recovery. This method improves the reliability and stability of the distribution network through real-time monitoring and data analysis.

Xingru et al. \cite{xingru2024} proposed a fault diagnosis and self-healing control model for power distribution network. The model combines an intelligent control mechanism to realize automatic fault detection and recovery to ensure the stable operation of the power system. 

Karamthulla et al. \cite{karamthulla2023} explore the paradigm of applying AI-driven self-healing systems in fault-tolerant platform engineering. Through a series of case studies, the application of AI technologies such as machine learning and neural networks in creating self-healing mechanisms is analyzed, and challenges such as scalability, adaptability, and robustness are discussed, as well as problems in practical implementation.

Walter \cite{walter2024} proposes a paradigm shift from a reactive security model to an active, autonomous security mechanism. The study discusses in detail the architecture, working principles, and deployment methods of self-healing AI systems, and provides an in-depth analysis of autonomous threat detection technologies, automatic remediation mechanisms, and challenges and ethical considerations.

\section{METHODOLOGIES}
\subsection{Fault detection}
To more precisely capture potential information in syslogs and error reports, we introduced time series embedding for data preprocessing. For time series data $X = \{x_1, x_2, \ldots, x_n\}$, we generate dynamic features via a long short-term memory network (LSTM), as shown in Equation~\eqref{eq:lstm}:
\begin{equation}
    h_t = \mathrm{LSTM}(h_{t-1}, x_t), \quad t = 1, 2, \ldots, n,
    \label{eq:lstm}
\end{equation}
where $h_t$ is the hidden state at time $t$, $x_t$ is the input feature at time $t$, and $n$ is the length of the input sequence.

Further, in order to enhance the expressive ability of the model, we jointly represent these time series data with the text feature $E_{\text{text}}$ generated by the natural language processing model (e.g., BERT), as shown in Equation~\eqref{eq:bert}:
\begin{equation}
    E_{\text{text}} = \text{BERT}(T_i), \quad i = 1, 2, \ldots, N,
    \label{eq:bert}
\end{equation}
where $T_i$ is the $i$-th log or report, and $E_{\text{text}}$ is the embedding vector generated by BERT.

In the fault detection process, we combine a support vector machine (SVM) with an autoencoder model. In the supervised learning part, we use the objective function of the SVM model as Equation~\eqref{eq:svm}:
\begin{equation}
    \min_{w, b, \boldsymbol{\xi}} \frac{1}{2} \lVert w \rVert^2 + C \sum_{i=1}^{n} \xi_i,
    \label{eq:svm}
\end{equation}
where $w$ is the weight vector of the classifier, $b$ is the bias term, $\xi_i$ is the relaxation variable, and $C$ is the penalty coefficient, which aims to balance the classification error and the complexity of the model.

In the unsupervised learning part, we use an autoencoder to perform anomaly detection on the data. To capture higher-order features, we introduce regularization terms into the autoencoder, such as Equation~\eqref{eq:ae_loss}:
\begin{equation}
    \mathcal{L}_{AE}(x, \hat{x}) = \lVert x - \hat{x} \rVert_2^2 + \lambda \lVert W \rVert_2^2,
    \label{eq:ae_loss}
\end{equation}
where $x$ is the input data, $\hat{x}$ is the reconstructed data, $W$ is the weight matrix of the autoencoder, and $\lambda$ is the regularization coefficient to ensure that the model does not overfit.

To further enhance the robustness of the model, we used a variational autoencoder (VAE) as a deep generative model to detect potential anomalous patterns. The variational lower bound (ELBO) loss function for VAE is given by Equation~\eqref{eq:vae_loss}:
\begin{equation}
    \mathcal{L}_{VAE} = \mathbb{E}_{q(z|x)}[\log p(x|z)] - D_{\mathrm{KL}}[q(z|x) \parallel p(z)],
    \label{eq:vae_loss}
\end{equation}
where $z$ is the latent variable, $q(z|x)$ is the variational posterior distribution, $p(z)$ is the prior distribution, and $D_{\mathrm{KL}}$ is the Kullback-Leibler divergence.

\subsection*{B. Fault self-healing mechanism}

In order to improve the accuracy of failure prediction, we use an LSTM-based time series model to predict the possible failures of the system at the next $t + k$ moment. The state update equation for the LSTM is Equation~\eqref{eq:lstm2}:
\begin{equation}
    h_t = \sigma(W_h h_{t-1} + W_x x_t + b_h),
    \label{eq:lstm2}
\end{equation}
where $h_t$ is the hidden state of moment $t$, $W_h$ and $W_x$ are the weight matrices, $x_t$ is the input feature, $b_h$ is the bias term, and $\sigma$ is the activation function.

In order to predict potential failures before they occur, we combine time series features with cloud system state features to further improve prediction capabilities through deep neural networks (DNNs). The loss function of the DNN is given by Equation~\eqref{eq:dnn_loss}:
\begin{equation}
    \mathcal{L}_{DNN} = \frac{1}{n} \sum_{i=1}^{n} \lVert y_i - \hat{y}_i \rVert^2,
    \label{eq:dnn_loss}
\end{equation}
where $y_i$ is the true label, $\hat{y}_i$ is the predicted value, and $n$ is the number of samples.

In terms of the self-healing mechanism, the system optimizes the decision-making process through reinforcement learning (RL), and the reward function is defined as Equation~\eqref{eq:reward}:
\begin{equation}
R(s_t, a_t) =
\begin{cases}
+1, & \textit{if the system recovers successfully} \\
   & \textit{after fault detection} \\
-1, & \textit{if the system fails to} \\
  & \textit{recover or causes additional damage} \\
0,  & \textit{otherwise}
\end{cases}
\label{eq:reward}
\end{equation}

where $s_t$ denotes the state of the system and $a_t$ denotes the action taken.

In order to optimize the overall system, we use a joint training method that combines supervised learning and unsupervised learning with joint optimization of fault classification and anomaly detection. The loss function is given in Equation~\eqref{eq:total_loss}:
\begin{equation}
\mathcal{L}_{\text{total}} = \mathcal{L}_{SVM} + \mathcal{L}_{AE} + \mathcal{L}_{VAE} + \mathcal{L}_{DNN} + \mathcal{L}_{RL}
\label{eq:total_loss}
\end{equation}
Among them, $\mathcal{L}_{SVM}$, $\mathcal{L}_{AE}$, $\mathcal{L}_{VAE}$, $\mathcal{L}_{DNN}$, and $\mathcal{L}_{RL}$ represent the loss functions of support vector machines, autoencoders, variational autoencoders, deep neural networks, and reinforcement learning, respectively.

The model is trained with parameter updates by a backpropagation algorithm, and the optimization goal is to minimize the total loss function, as shown in Equation~\eqref{eq:update}:
\begin{equation}
\theta_{t+1} = \theta_t - \eta \nabla_{\theta} \mathcal{L}_{\text{total}}(\theta_t),
\label{eq:update}
\end{equation}
where $\theta_t$ is the model parameter, $\eta$ is the learning rate, and $\nabla_{\theta} \mathcal{L}_{\text{total}}(\theta_t)$ is the gradient of the total loss function.

\section{EXPERIMENTS}
\subsection{Experimental setup}
The experiment used the Verified Telemetry (VT) SDK dataset developed by Microsoft Research, which is designed for sensor failure detection in the Internet of Things (IoT) environment. Through sensor fingerprint technology, the dataset supports multiple sensor types, enabling real-time detection of device failures and reducing reliance on cloud resources. The VT SDK has been proven in practical applications such as agricultural monitoring and environmental monitoring to ensure the accuracy of fault detection and the reliability of the system.

We select four detection models and methods to compare including:
\begin{itemize}
    \item Transformers for Fault Detection (Transformers-FD) has been introduced into fault detection tasks due to its successful application in natural language processing, especially when working with long-term series data. Transformers-FD is excellent at detecting and predicting faults in cloud computing and complex systems because it can efficiently capture long-term dependencies in data by using the self-attention mechanism.
    \item Graph Neural Networks (GNNs) for Fault Detection has shown great potential in the field of fault detection, especially for troubleshooting in complex and distributed systems, by disseminating information through relationships between nodes
    \item Deep Reinforcement Learning (DRL) for Autonomous Self-Healing combines the advantages of deep learning with reinforcement learning to excel in automated decision-making and control tasks. Recently, DRLs have been introduced into self-healing systems to autonomously learn to optimize control strategies by interacting with the environment to reduce failure recovery time and improve system adaptability. 
    \item BERT-based Anomaly Detection for Cloud Systems (BERT) has achieved remarkable results in the field of natural language processing (BERT). Recent research has applied BERT to anomaly detection in cloud computing systems, particularly for the processing of unstructured data from system logs and error reports.
\end{itemize}

\subsection{Experimental analysis}
Accuracy reflects the proportion of predictions that the model correctly predicts out of all predictions. For fault detection tasks, a high accuracy rate means that most faults are correctly identified. 

\begin{figure}[h!]
  \centering
    \includegraphics[width=0.9\linewidth, height=0.45\linewidth]{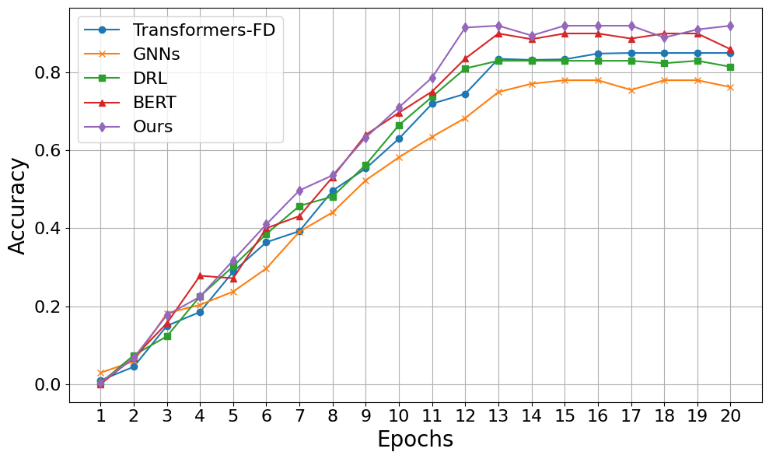}
    \caption{Accuracy Comparison of Fault Detection Methods.}
  \label{fig:bar}
\end{figure}

The proposed model Ours in Figure 1 showed excellent performance in the fault detection task, and the accuracy rate increased rapidly during the training process and finally reached about $92\%$, and remained stable in the subsequent epochs. This shows that our method quickly grasps the key features in the data through efficient learning in the early stage, and successfully achieves convergence in the later stage to achieve high accuracy. 

Compared to other comparison methods, Transformers-FD and BERT also performed very close at convergence, with an accuracy rate of nearly $85\%$, but their training fluctuations were large and not as stable as our method. GNNs and DRLs were relatively smooth in the early stages of training, but their final accuracy and convergence were slightly lower in comparison, reaching about $78\%$ and $83\%$, respectively.

Failure recovery time is a key measure of the importance of a self-healing system, referring to the time it takes to go from failure detection to completion of recovery operations. Figure 2 compares the failure recovery time of different methods in different feature dimensions, i.e., model complexity. 

\begin{figure}[h!]
  \centering
    \includegraphics[width=0.9\linewidth, height=0.45\linewidth]{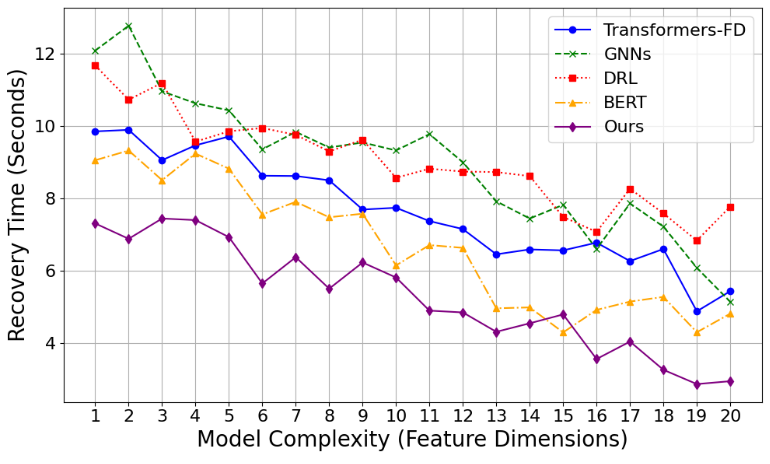}
    \caption{Fault Recovery Time Comparison with Model Complexity.}
  \label{fig:bar}
\end{figure}

With the increase of feature dimension, the recovery time of each model shows different trends. The recovery time of Transformers-FD and GNNs was long at the beginning, but gradually decreased and stabilized with the increase of feature dimension. In contrast, DRL and BERT showed faster recovery in the initial stage, but the recovery time fluctuated slightly as the feature dimension increased, showing the adaptability and stability of the model under high-dimensional data.

System stability measures the performance of the system in response to different fault situations during operation. Figure 3 shows the robustness of different approaches when dealing with complex system failures.

\begin{figure}[h!]
  \centering
    \includegraphics[width=0.9\linewidth, height=0.45\linewidth]{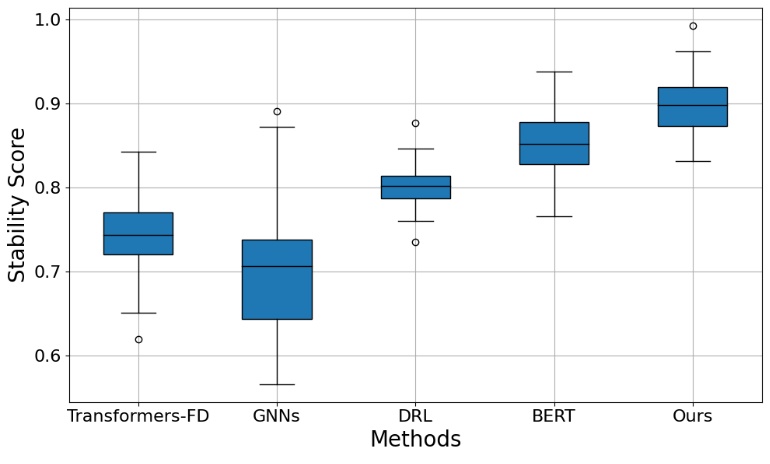}
    \caption{System Stability Comparison}
  \label{fig:bar}
\end{figure}

The Ours method has the highest stability score and the smallest range of data fluctuations, indicating that the method has consistently maintained high stability across multiple experiments. In contrast, BERT is slightly less stable, but still exhibits higher stability, especially in reducing volatility.

\section{CONCLUSION}
In conclusion, the proposed Ours method performs well in fault detection and self-healing tasks, with the best accuracy, fast recovery time and excellent system stability. Compared to other advanced methods, Ours has demonstrated greater robustness and reliability. Future research can further optimize the model to cope with more complex cloud environments, and explore how to combine more real-time data and heterogeneous systems to improve the adaptability and efficiency of the model.

\renewcommand{\bibfont}{\footnotesize}

\footnotesize{
\bibliographystyle{IEEEtran}
\bibliography{main}
}

\end{document}